\documentclass[authoryear,preprint,12pt]{style_}

\usepackage{graphicx}

\usepackage{amssymb}

\begin{document}

\begin{frontmatter}



\title{Phase-fitted Discrete Lagrangian Integrators}


\author[uop]{O.T. Kosmas}
\ead{odykosm@uop.gr}

\author[uop]{D.S. Vlachos\corref{cor}}
\ead{dvlachos@uop.gr}

\cortext[cor]{Corresponding author}
\address[uop]{Department of Computer Science and Technology,\\
Faculty of Sciences and Technology, University of Peloponnese\\
GR-22 100 Tripolis, Terma Karaiskaki, GREECE}

\begin{abstract}
Phase fitting has been extensively used during the last years to improve the behaviour of numerical integrators on oscillatory problems. In this work, the benefits of the phase fitting technique are embedded in discrete Lagrangian integrators. The results show improved accuracy and total energy behaviour in Hamiltonian systems. Numerical tests on the long term integration ($10^5$ periods) of the 2-body problem with eccentricity even up to 0.95 show the efficiency of the proposed approach. Finally, based on a geometrical evaluation of the frequency of the problem, a new technique for adaptive error control is presented.\end{abstract}


\end{frontmatter}

\section{Introduction}
\label{section_intro}
In the field of numerical integration, methods specially tuned on oscillating functions, are of great practical importance. Such methods are needed in various branches of natural sciences, particularly in physics, since a lot of physical phenomena exhibit a pronounced oscillatory behaviour. For a review of such methods see \cite{simos_Book_CMATV1_RCS_00,ixaru_CPC_100_56_70,vandenberghe_CPC_123_7_15,vandenberghe_CPC_150_346_357,ixaru_CPC_150_116_128,vandaele_APNUM_57_415_435} and references there in as well as the book \cite{ixaru_Book_EF_KAP_04}.

For problems having highly oscillatory solutions, standard methods with unspecialised use can require a huge number of steps to track the oscillations. One way to obtain a more efficient integration process is to construct numerical methods with an increased algebraic order, although the simple implementation of high algebraic order methods may cause several problems (for example, the existence of parasitic solutions \cite{quinlan_arxiv_astro_ph_9901136}). On the other hand, there are some special techniques for optimising numerical methods. Trigonometrical fitting and phase-fitting are some of them, producing methods with variable coefficients, which depend on $v = \omega h$, where $\omega$ is the dominant frequency of the problem and $h$ is the step length of integration. This technique is known as exponential (or trigonometric if $\mu=i\omega$) fitting and has a long history \cite{gautschi_NM_3_381_61}, \cite{lyche_NM_19_65_72}. An important property of exponential fitted algorithms is that they tend to the classical ones when the involved frequencies tend to zero, a fact which allows to say that exponential fitting represents a natural extension of the classical polynomial fitting. The examination of the convergence of exponential fitted multistep methods is included in Lyche’s theory \cite{lyche_NM_19_65_72}. The general theory is presented in detail in \cite{ixaru_Book_EF_KAP_04}. Furthermore, considering the accuracy of a method when solving oscillatory problems, it is more appropriate to work with the phase-lag, rather than its usual primary local truncation error. We mention the pioneering paper of Brusa and Nigro \cite{brusa_IJNME_15_685_80}, in which the phase-lag property was introduced. This is actually another type of a truncation error, i.e. the angle between the analytical solution and the numerical solution. A significant application of the phase or exponential fitting is on the construction of symplectic methods for oscillatory problems encounterd in physics and chemistry (\cite{monovasilis_JMC_37_3263_05,monovasilis_JMC_40_3257_06}).

Another approach to oscillatory and especially Hamiltonian systems is the theory of discrete variational mechanics, which was set up in the 1960s \cite{jordan_JEL_17_697_64,cadzow_IJC_11_393_70,logan_AM_9_210_73} and then it was proposed in the optimal control literature. It then motivated a lot of authors and soon the discrete Euler-Lagrange equations were formulated and the first integrators in the discrete calculus of variation and further the multi-freedom and higher-order problems were studied. Afterwards, the canonical structure and symmetries for discrete systems were obtained, and Noether's theorem to the discrete case was extended \cite{maeda_MJ_25_405_80,maeda_MJ_26_85_81}. Finally, the time as a discrete dynamical variable was regarded \cite{lee_PLB_122_217_83}. A detailed description of the essential properties of variational integrators can be found in \cite{marsden_AN_10_357_01,marsden_CMP_199_351_98,lee_PLB_122_217_83}. One of the most important properties of variational integrators is that since the discrete Lagrangian is an approximation of a continuous Lagrangian function, the obtained numerical integrator inherits some of the geometric properties of the continuous Lagrangian (such as symplecticity, momentum preservation).

In the present work, the benefits of the two approach are combined in order to construct discrete Lagrangian integrators with phase fitting. To obtain this, we have adopted a test Lagrangian problem (similar to the test ODE in the phase fitting) which is the harmonic oscillator with given frequency $\omega$. Then, we construct discrete variational schemes that solve exactly the test Lagrangian. The application of the method to a general Lagrangian needs the determination of the frequency $\omega$ at every step of the integration. The method is applied to the 2-body problem with eccentricity up to 0.95. The results exhibit an improved behaviour of the calculated solutions, especially in the case of total energy of the integrated systems.

\section{Discrete Variational Mechanics}
\label{section_DVM}
The well known least action principle of the continuous Lagrange - Hamilton Dynamics can be used as a guiding principle to derive discrete integrators. Following the steps of the derivation of Euler-Lagrange equations in the continuous time Lagrangian dynamics, one can derive the discrete time Euler-Lagrange equations. For this purpose, one considers positions $q_{0}$ and $q_{1}$ and a time step $h\in{R}$, in order to replace the parameters of position $q$ and velocity $\dot{q}$ in the continuous time Lagrangian $L (q, \dot{q}, t)$. Then, by considering the variable $h$ as a very small (positive) number, the positions $q_{0}$ and $q_{1}$ could be thought of as being two points on a curve (trajectory of the mechanical system) at time $h$ apart. Under these assumptions, the following approximations hold: $$q_{0}\approx q(0)\, , \qquad \qquad q_{1}\approx q(h) \, ,$$ and a function $L_{d}(q_{0},q_{1},h)$ could be defined known as a discrete Lagrangian function.

Many authors assume such functions to approximate the action integral along the curve segment between $q_{0}$ and $q_{1}$, i.e.
\begin{equation}
 L_{d}(q_{0},q_{1},h)=\int_{0}^{h}L(q,\dot{q},t) dt
\end{equation}
Furthermore, one may consider the very simple approximation for this integral given on the basis of the rectangle rule described in \cite{marsden_AN_10_357_01}. According to this rule, the integral $\int_{0}^{T}{Ldt}$ could be approximated by the product of the time-interval ${h}$ times the value of the integrand $L$ obtained with the velocity $\dot{q}$ replaced by the approximation $(q_{1}-q_{0})/h$:  The next step is to consider a discrete curve defined by the set of points $\{q_{k}\} _{k=0}^{N}$, and calculate the discrete action along this sequence by summing the discrete Lagrangian of the form $L_{d}(q_{k},q_{k+1},h)$ defined for each adjacent pair of points $(q_{k}$, $q_{k+1})$. 

Following the case of the continuous dynamics, we compute variations of this action sum with the boundary points $q_{0}$ and $q_{N}$ held fixed. Briefly, discretization of the action functional leads to the concept of an action sum 
\begin{equation}
S_{d}(\gamma_{d})=\sum_{k=1}^{n-1}L_{d}(q_{k-1},q_{k}),
\qquad \gamma_{d}=(q_{0},...,q_{n-1})\in Q^{n}
\end{equation}
where $L_{d} : Q \times Q \rightarrow R$ is an approximation of L called the discrete Lagrangian. Hence, in the discrete setting the correspondence to the velocity phase space $TQ$ is $Q \times Q$. An intuitive motivation for this is that two points close to each other correspond approximately to the same information as one point and a velocity vector. The discrete Hamilton's principle states that if $\gamma_{d}$ is a motion of the discrete mechanical system then it extremizes the action sum, i. e., $\delta S_{d}=0$. By differentiation and rearranging of the terms and having in mind that both $q_0$ and $q_N$ are fixed, the discrete Euler-Lagrange (DEL) equation is obtained:
\begin{equation}
\label{equ_DEL}
D_{2}L_{d}(q_{k-1},q_{k},h)+D_{1}L_{d}(q_{k},q_{k+1},h)=0
\end{equation}
where the notation $D_{i}L_{d}$ indicates the slot derivative with respect to the argument of $L_{d}$.

In a position-momentum form the discrete Euler-Lagrange equations (\ref{equ_DEL}) can be defined by the equations below 
\begin{eqnarray}
\label{equ_DHP}
\nonumber p_{k}&=&-D_{1}L_{d}(q_{k},q_{k+1},h) \\
p_{k+1}&=&D_{2}L_{d}(q_{k},q_{k+1},h)
\end{eqnarray}

\section{Phase-fitted Discrete Lagrangian Integrators}
Summarising the phase fitting technique, we consider for simplicity only first order differential equations, although the same results can be easily obtained for second order equations too. Consider the test problem 
\begin{equation}
\frac{dy(t)}{dt}=i\omega _0 y(t),\;y(0)=1 
\label{equ_test_equ}
\end{equation}
with exact solution 
\begin{equation}
y(t)=e^{i\omega _0 t}
\end{equation}
where $\omega _0$ is a non-negative real value. Let $\hat{\Phi}(h)$ be a numerical map which when it is applied to a set of known past values, it produces a numerical estimation of $y(t+h)$. If we assume that all past values are known exactly, then the numerical estimation $\hat{y}(t+h)$ of $y(t+h)$ will be
\begin{equation}
\hat{y}(t+h)=\alpha (\omega _0 h) \cdot e^{i(\omega _0 t+\phi (\omega _0 h) )}
\end{equation}
while the exact solution is $e^{i (\omega _0 t +\omega _0 h)}$. Then the ratio of the estimated to the exact solution is
\begin{equation}
L=\frac{\hat{y}(t+h)}{y(t+h)}=\alpha (\omega _0 h)e^{-i(\omega _0 h-\phi (\omega
_0 h) ) }
\label{equ_def_PL}
\end{equation}
In the above equation (\ref{equ_def_PL}), the term $\alpha (\omega _0 h)$ is called the \textit{amplification error}, while the term $l(\omega _0 h)=\omega
_0 h-\phi(\omega _0 h)$ is called the \textit{phase lag} of the numerical map. In the case that $\alpha (\omega _0 h)=1$ and $l(\omega _0 h)=0$, we say that
the numerical map $\Phi(h)$ is \textit{exponentially fitted} at the frequency $\omega _0$ and at the step size $h$. The technique of phase fitting can now be considered as the vanishing or minimisation of the phase lag.

Consider now the discrete Lagrangian $L_d(q_k,q_{k+1},h)$ ($q_k$ corresponds to time $t_k$ and $q_{k+1}$ to time $t_{k+1}=t_k+h$) and a set of $s$ intermediate points $q^j$ with $q^j=q\left(t_k+c^jh\right)$. The role of the number of intermidiate points will be discussed later. Assuming that $c^1=0$ and $c^s=1$ we always have $q^1=q_k$ and $q^s=q_{k+1}$. Then we can approximate $L_d$ with the quadrature
\begin{equation}
 L_d(q_k,q_{k+1},h)=h\sum_{j=1}^s w_j\cdot L\left( q(t_k+c^jh),\dot{q}(t_k+c^jh),c^jh \right)
\end{equation}
For maximal algebraic order it is easily proved that the following conditions must hold:
\begin{equation}
 \sum_{j=1}^s w_j\left( c^j \right) ^l=\frac{1}{l+1}\;,\;l=0,1,..
\end{equation}
Then, we can approximate intermediate points and their derivatives with
\begin{equation}
\label{equ_inter}
 \begin{array}{l}
  q^j=b^jq_k+\bar{b}^jq_{k+1} \\
  \dot{q}^j=\frac{1}{h}\left( B^jq_k+\bar{B}^jq_{k+1}\right)
 \end{array}
\end{equation}
Consider now the test Lagrangian (harmonic oscillator) similar to the test equation (\ref{equ_test_equ})
\begin{equation}
\label{equ_Ltest}
 L_t=\frac{1}{2}\dot{q}^2-\frac{1}{2}\omega ^2 q^2
\end{equation}
Then, applying the above assumptions in Eq. (\ref{equ_DEL}) we get
\begin{equation}
 q_{k+1}=\frac{\sum_{j=1}^sw_j\left( \left(B^j\right)^2+\left(\bar{B}^j\right)^2-u^2\left(\left(b^j\right)^2+\left(\bar{b}^j\right)^2\right)\right)}{\sum_{j=1}^sw_j\left(u^2b^j\bar{b}^j-B^j\bar{B}^j\right)}q_k-q_{k-1}
\end{equation}
where $u=\omega h$. Since the exact solution of Eq. (\ref{equ_Ltest}) is
\begin{equation}
 q(t)=Ae^{i\omega t}+Be^{-i\omega t}
\end{equation}
and we want to force our method to solve exactly Eq. (\ref{equ_DEL}), we get
\begin{eqnarray}
\nonumber   b^j&=&cos \left( c^ju \right)-\frac{cosu}{sinu}sin \left( c^ju \right) \\
\nonumber \bar{b}^j&=&\frac{ sin \left( c^ju \right) }{sinu} \\
\nonumber B^j&=&-usin\left( c^ju \right)-u\frac{cosu}{sinu}cos\left( c^ju \right)\\
\bar{B}^j&=&u\frac{cos\left( c^ju \right) }{sinu}
\end{eqnarray}

\section{Frequency Evaluation and Error Control}
\label{sec_FE}
\begin{figure}
\center
\includegraphics[scale=0.3]{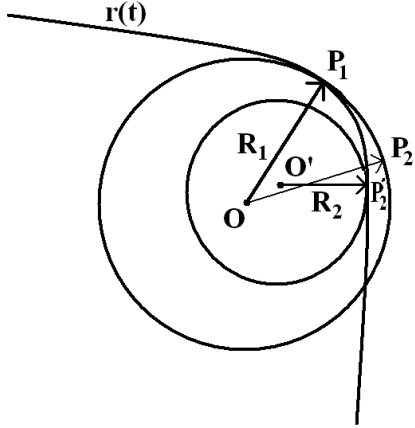}
\caption{The oscullating circles at two neighbour points $P_1$ and $P_2'$ are shown. The first circle has its centre at point $O$ and radius $R1$ and the second its centre at point $O'$ with radius $R2$. If $P_2$ is the estimated point and the time step $h$ is small enough, then the distance between $P_2$ and $P_2'$ is of the order of the absolute difference of the two radii minus the distance between $O$ and $O'$. }
\label{fig_err_control}
\end{figure}

The final step to our method is to evaluate the frequency of the problem. For this purpose we use the curvature of the solution and the concept of the oscullating circle. Consider a planar curve $C$ and a point $P$ on this curve. If $\textbf{r}(t)$ is a parametrised representation of the curve, then we define the curvature at a point $P$ as
\begin{equation}
 k(t)=\frac{\dot{\textbf{r}}(t)\times\ddot{\textbf{r}}(t)}{|\dot{\textbf{r}}(t)|^3}
\end{equation}
Then, there is a circle with radius
\begin{equation}
 R=\frac{1}{|k(t)|}
\end{equation}
which locally approximates the curve at point $\textbf{r}(t)$ and is called the oscullating circle. Since, the velocity of a point running on top of the curve $C$ is $|\dot{\textbf{r}}(t)|$, at a small time step $h$ the point will rotate an angle equal to
\begin{equation}
 \theta=\frac{|\dot{\textbf{r}}(t)|}{R}h=\frac{|\dot{\textbf{r}}\times\ddot{\textbf{r}}|}{|\dot{\textbf{r}}|^2}h
\end{equation}
leading us to a frequency selection
\begin{equation}
\label{equ_freq}
 \omega=\frac{|\dot{\textbf{r}}\times\ddot{\textbf{r}}|}{|\dot{\textbf{r}}|^2}
\end{equation}
Consider now the regular parametrisation of the curve $\textbf{r}$ (the one that uses the curve length as the free parameter). Let $s$ the curve length. Then the curve $\textbf{u}$ of the centres of the oscullating circles is given by
\begin{equation}
 \textbf{u}(s)=\textbf{r}(s)+\frac{1}{k(s)}\textbf{H}(s)
\end{equation}
where $\textbf{H}(s)$ is the first normal vector of the curve $\textbf{r}$ at point $\textbf{r}(s)$. Then,
\begin{equation}
 \frac{d\textbf{u}(s)}{ds}=-\frac{dk(s)/ds}{k(s)^2}\textbf{H}(s)
\end{equation}
which means that the centre of the oscullating circle is moving to a direction normal to the curve $\textbf{r}$ with velocity
\begin{equation}
 v_o=\left|\frac{dk(s)/ds}{k(s)^2}\right|
\end{equation}
Assuming now a small displacement on curve $\textbf{r}$, it can be easily proved that the distance that is covered by the centre of the oscullating circle is smaller than the difference of their radius which means that the one circle is entirely inside or outside of the other (depending on the variation of the curvature). Thus, the error in position is bounded by
\begin{equation}
\label{equ_error}
 \left|\left|\frac{1}{k_1}-\frac{1}{k_2}\right|-v_0h\right|
\end{equation}
where $k_1,k_2$ are the curvatures of the curve $\textbf{r}$ at the adjacent points and $h$ is the time step. Figure (\ref{fig_err_control}) depicts this result. Using Eq. (\ref{equ_error}) we can adaptively control the time step of the integration, keeping the local truncation error within desired bounds.

\section{Numerical Test for the 2-body Problem}
We now turn to the study of two objects interacting through a central force. The most famous example of this type, is the Kepler
problem (also called the two-body problem) that describes the motion of two bodies which attract each other. In the solar system the gravitational interaction between two bodies leads to the elliptic orbits of planets and the hyperbolic orbits of comets.

\begin{figure}
\center
\includegraphics[scale=0.3]{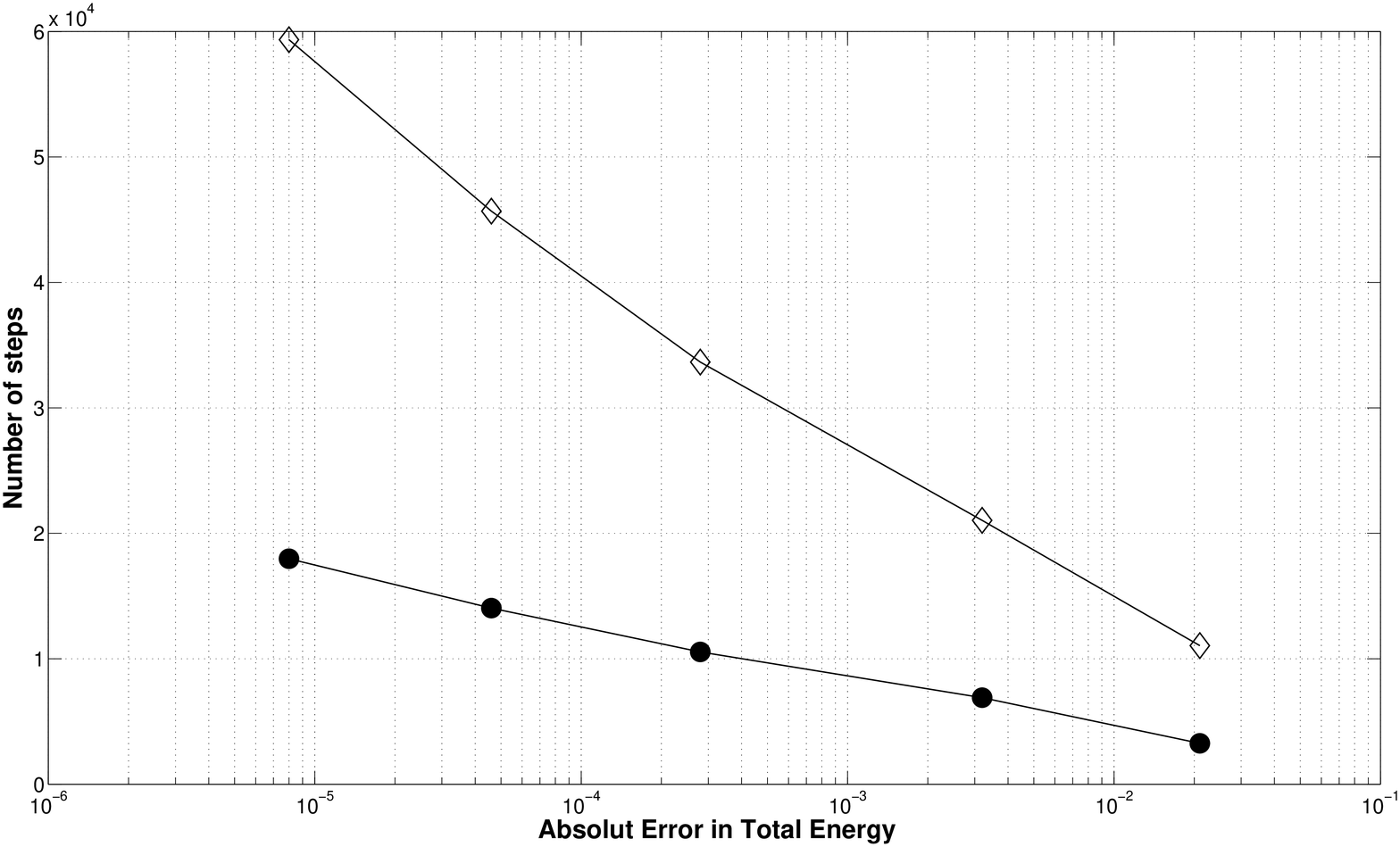}
\caption{The number of integration steps needed to obtain the same accuracy in total energy for the proposed method ($\bullet$) and for the method described in \cite{marsden_AN_10_357_01} ($\Diamond$).}
\label{fig_res1}
\end{figure}

If we choose one of the bodies as the centre of our coordinate system, the motion will stay in a plane. Denoting the position of
the second body by $\textbf{q}=(q_{1},q_{2})^{T}$, the Lagrangian of the system takes the form (assuming masses and gravitational constant equal to 1)
\begin{equation}
L(\textbf{q},\dot{\textbf{q}},t)=\frac{1}{2}\dot{\textbf{q}}^T \dot{\textbf{q}}+\frac{1}{|\textbf{q}|}
\end{equation}
The initial conditions are taken
\begin{equation}
 \textbf{q}=(1-\epsilon,0)^T\;,\;\dot{\textbf{q}}=\left(0,\sqrt{\frac{1+\epsilon}{1-\epsilon}}\right)^T
\end{equation}
where $\epsilon$ is the eccentricity of the orbit. In order to check the efficiency of the proposed algorithm, we shall consider only high eccentricity ($\epsilon=0.95$). Figure (\ref{fig_res1}) compares the proposed method with the method described in \cite{marsden_AN_10_357_01} (the two methods have the same algebraic order). The results here are obtained as follows: First, the phase fitted method is applied for one period and for a given tolerance in position calculation. Then, the energy tolerance is calculated and the method of \cite{marsden_AN_10_357_01} is applied using a variable step length in order to obtain the same energy tolerance. It is clear that the phase fitting decreases close to one third the number of integration steps to obtain the same accuracy. 

\begin{figure}
\center
\includegraphics[scale=0.3]{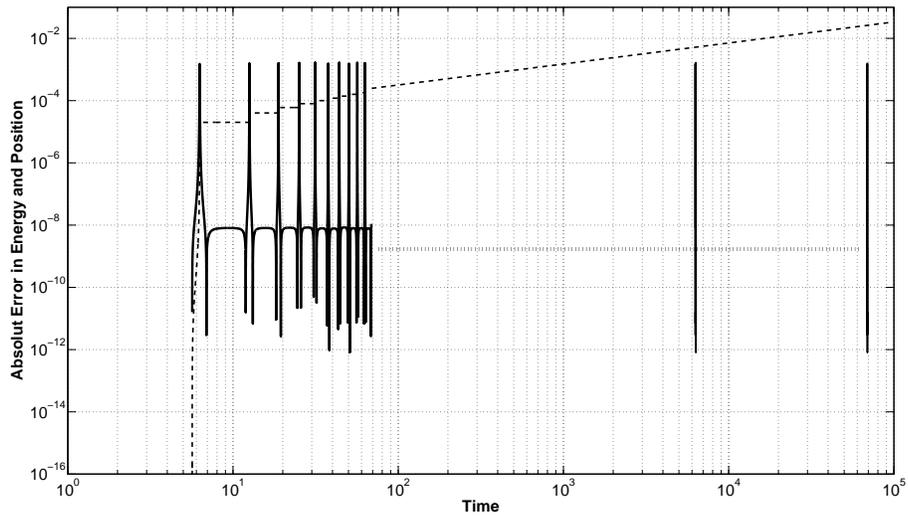}
\caption{The total energy (solid line) end the error in position (dashed line) as a function of time for eccentricity 0.95 and for $10^5$ periods.}
\label{fig_res2}
\end{figure}

\begin{figure}
\center
\includegraphics[scale=0.3]{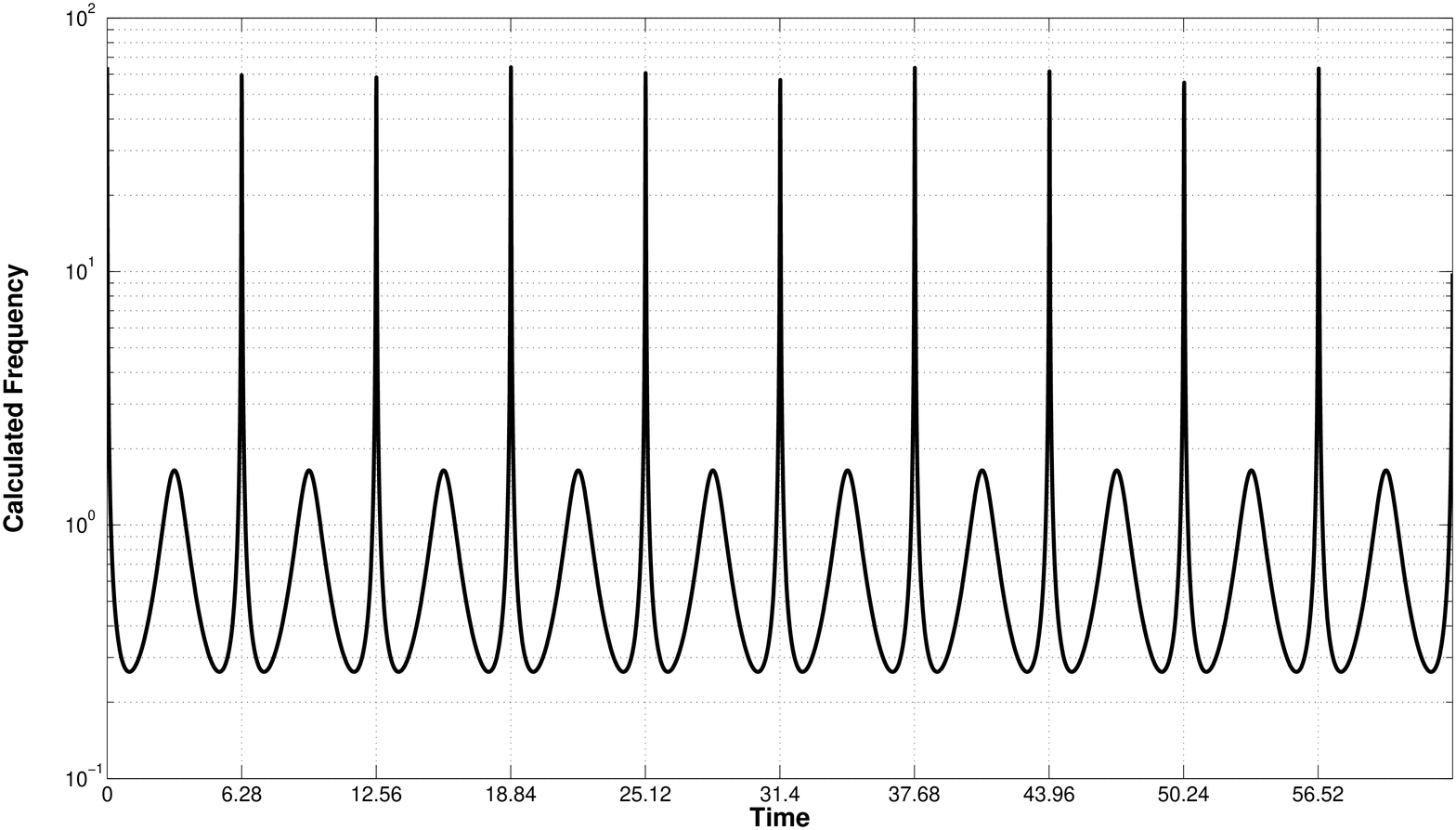}
\caption{The calculated frequency of the problem is shown for the $10$ first periods.}
\label{fig_res3}
\end{figure}

Finally, in figure \ref{fig_res2} the total energy is plotted as a function of time for $10^5$ periods as well as the error in position (the distance between the calculated and the exact points). It is clear that the method keeps both energy and position error in stable limits although there is a relative increased error in energy at the perihelion. This can be explained by considering the calculated from equation Eq. (\ref{equ_freq}) frequency of the problem(see Fig. \ref{fig_res3}). The frequency is smooth enough almost everywhere except at the perihelion where it changes rapidly. This means that the assumption that the frequency is constant during an integration step, applies everywhere except at the perihelion and this is the reason for the observed increase in the total energy error. This undesirable effect can be handled by increasing the algebraic order of the method.

\section{Increasing the algebraic order}
Until now, he have not mention anything about the role of the number of intermediate points used to calculate the discrete Lagrangian. This can be used to increase the algebraic order of the method. Consider Eq. (\ref{equ_inter}) modified as
\begin{equation}
 \begin{array}{l}
  q^j=b^jq_k+\bar{b}^jq_{k+1} +\delta q^j\\
  \dot{q}^j=\frac{1}{h}\left( B^jq_k+\bar{B}^jq_{k+1}\right)+\delta \dot{q}^j
 \end{array}
\end{equation}
where the corrections $\delta q^j$ are free parameters and the corresponding corrections $\delta \dot{q}^j$ for the derivatives cane be calculated
\begin{equation}
 \delta \dot{q}^j=\frac{1}{h}\sum_{k=1}^sa^j_k\delta q^k
\end{equation}
where the coefficients $a^j_k$ can be easily calculated for maximal algebraic order as
\begin{eqnarray}
\nonumber \sum_{j=1}^sa^k_j&=&0 \\
\sum_{j=1}^{s}a^k_j\left(c^j\right)^n&=&n\left(c^k\right) ^{n-1}\;,\;n=1,2,...
\end{eqnarray}
Now, the discrete Lagrangian, beyond $(q_k,q_{k+1})$ depends also on $\delta q^j$. The system of Eq. (\ref{equ_DHP}) now is enriched with the equations (since we want the discrete Lagrangian to be stationary)
\begin{equation}
 \frac{\partial L_d(q_k,q_{k+1},h)}{\partial \delta q^j}=0\;,\;j=1,2,...,s
\end{equation}
This technique is similar to those described in \cite{leok_arXiv_math_0508360_05} and \cite{kharevych_AGM_SIGGRAPH_06}.

\section{Conclusions}
It has been shown in this work, that the technique of phase fitting, when it is embedded in discrete Lagrangian integrators, improves the accuracy and the energy behaviour of the numerical method. Following the classical application of the phase fitting technique, the discrete Lagrangian integrator is forced to solve exactly the test Lagrangian of harmonic oscillator with a given self frequency. The coefficients of the resulting integrator, depend on the frequency of the problem at each integration step. Geometrical consideration lead us to a new frequency evaluation depending on the curvature of the solution and on the principle of the oscullating circle. Furthermore, the resulting analysis gave us a new error control technique. Application of the new method to the well known 2-body problem decreased the number of integration steps needed to obtain the same accuracy with the classical discrete Lagrangian of the same algebraic order to less than one third in high eccentricity equal to 0.95. Moreover, the new method exhibits improved energy behaviour for long term integration ($10^5$ periods of the 2-body problem with eccentricity equal to 0.95). Finally, a simple method is proposed to increase the algebraic order of the new method.

\section{Acknowledgement}
This paper is part of the 03ED51 research project, implemented within the framework of the "\emph{Reinforcement Programme of Human Research Manpower}" (\textbf{PENED}) and co-financed by National and Community Funds (25\% from the Greek Ministry of Development-General Secretariat of Research and Technology and 75\% from E.U.-European Social Fund).




\bibliographystyle{elsarticle-harv}

\end{document}